\documentclass[12pt]{article}

\usepackage{graphicx }
\usepackage{caption}
\usepackage{amsmath}
\usepackage{pdfpages}
\usepackage{makecell}
\usepackage{listings}
\usepackage{subfig}
\usepackage{wrapfig}
\usepackage{float}
\usepackage{rotating}
\usepackage{lscape}
\usepackage{authblk}
\usepackage[round]{natbib}

\title{The Effect of Multiple Imputation of Routine Pathology Variables on Laboratory Diagnosis of Hepatitis C Infection}
\author{Nidhi Menon\footnote{Corresponding author at: National Centre for Epidemiology \& Population Health, Australian National University, Canberra, Australia\\
email address: nidhi.menon@anu.edu.au } , Brett A. Lidbury \& Alice Richardson }
\affil{National Centre for Epidemiology \& Population Health, Australian National University, Canberra, Australia}

\date{}
			
\begin{document}
\maketitle
\begin{abstract}
\textbf{Background:} Pathology tests are central to modern healthcare in terms of diagnosis and patient management. Aggregated pathology results provide opportunities for research into fundamental and applied questions in health and medicine, but data analytic challenges appear since test profiles vary between medical practitioners, resulting in missing data. In this study we provide an analytical investigation of the laboratory diagnosis of Hepatitis C (HCV) infection and focus on how to maximize the predictive value of routine pathology data. We recommend using the Influx - Outflux measures to help construct the imputation model when using multiple imputation. \\
\\
\textbf{Methods:} Data from 14,320 community-patients aged 15 - 100 years were accessed via ACT Pathology (The Canberra Hospital, Australia). Influx and Outflux were calculated to identify which variables were potentially powerful predictors of missing values. Available Case analysis and Multiple Imputation were used to accommodate missing values in the dataset. Logistic regression model and stepwise selection method were used for analysing the imputed datasets. The predictive power of all methods was compared. \\
\\
\textbf{Results:} The predictive power of the models on multiply imputed data was similar to the power of the models based on complete data. The advantage of multiply imputed data was that it allowed for the inclusion of all the completed variables in the logistic models, thus identifying a broader selection of test results that could lead to the enhanced laboratory prediction of HCV.\\
\\
\textbf{Conclusions}: Multiple imputation is an important statistical resource allowing all individuals in a study to contribute whatever data they have supplied to the analysis. MI in combination with the values of Influx and Outflux identifies potential predictors of HepC infection. Variables age, gender and alanine aminotransferase have been shown to be strong laboratory predictors of HCV infection.\\
\\
\textbf{Keywords: }Hepatitis; logistic regression; multiple imputation.\\
\\
\textbf{Highlights:}
\begin{itemize}
\item  Multiple imputation of missing data is a well established technique in statistics and contribute to improving laboratory diagnosis. 
\item Laboratory diagnosis of Hepatitis C infection can be achieved using partially observed routine pathology data through multiple imputation.
\item Areas under ROC curve achieved in our analysis were around 70\% 
\item 	Multiple imputation in combination with the Influx - Outflux allows novel variables to contribute to the prediction models.
\item Age, gender and alanine aminotransferase are strong laboratory predictors of HCV infection.
\end{itemize}
\end{abstract}
\pagebreak

\section{Introduction}
Pathology laboratories generate large quantities of data representing human function, including analyses of blood chemistry and cells, infections, genetics and more. These data represent an under-utilized research resource, and while there are drawbacks \citep{1richardson2016clinical}, they introduce a rich informatics and statistical substrate to assist clinical decision making. They also support the interrogation of databases to assist answering research problems; for example, such databases have been successfully mined for enhanced laboratory prediction of infectious diseases \citep{2richardson2013infection, 3richardson2017enhancement}. Given the observational nature of the data collected, one drawback is that not every test is conducted for every patient. Therefore, pathology databases contain many missing values that potentially dilute the value of the data base.\\
\\
 Missingness can be one of three types: missing completely at random (MCAR), missing at random (MAR) and missing not at random (MNAR). Data are MCAR if the probability of missingness is not related to the value of the observation or any other variables in the data. For most datasets, the assumption of MCAR is unlikely to be satisfied unless the data is missing due to the design employed. If the missingness depends on the value of another variable in the dataset, then the data is said to be MAR. MCAR is a special case of MAR, thus if the data is MCAR, they are also MAR.  If the MAR assumption is violated, then the data is said to be MNAR \citep{4rubright2014simulation}.\\
 \\
Missing data mechanisms are important since older methods of handling missing data (e.g. available case analysis) assume MCAR missingness while modern techniques to handle missing data such as maximum likelihood estimation (MLE) and multiple imputation assume only MAR. 

\subsection{Hepatitis C Virus}
Hepatitis virus is a world-wide health concern, with an estimated 1.45 million deaths attributable to the virus globally in 2013 \citep{5stanaway2016global}.  More recent modelling \citep{6blach2017global} has also shown that the global prevalence of Hepatitis C virus (HCV) has reached an estimated 71.1 million infections. HCV affects the liver and can cause permanent and ultimately fatal liver cancer \citep{7el2012epidemiology}.   HCV causes the inflammation of the liver and over time this can lead to liver diseases like cirrhosis and fibrosis. HCV is a blood borne virus and in developed countries it is often transmitted through risky behaviour such as sharing needles. Indeed, in Australia in 2007 \citep{8razali2007modelling}, 90\% of the new and 80\% of the existing infections are transmitted in this way. \\
\\
While there are advances in the treatment of people with HCV, the human costs of HCV, in terms of reduction of quality of life and well being and through occupational and social discrimination and isolation, remain significant \citep{9vietri2013burden}. Guidelines for the care and treatment of HCV have recently been updated \citep{10world2018guidelines}.  The financial costs of the virus for medical and hospital care, lost productivity, and the need for social support are also an increasing proportion of healthcare expenditure in Australia.  While the diagnosis of HCV is most confidently made through the immunoassay \citep{11european2014easl}, having a powerful predictive model will help focus on early detection of HCV infection particularly for cases where the specific immunoassay has not been ordered.

\subsection{Multiple imputation}
Analysis of multivariate data is often negatively affected by incompleteness. Reduced statistical power, increase in standard errors, complications in data handling and analysis, and introduction of bias are some of the common problems associated with missing data \citep{12horton2001multiple}.  \\
\\
Several methods have been proposed over the years to handle the problem of missing data, for example single imputation using means or medians \citep{13little2002statistical} and the EM algorithm \citep{14dempster1977maximum}. While single imputation techniques are easy to implement, they treat the missing values as if they were known, thus eliminating any uncertainty that missing values bring to the dataset. They do not preserve the inherent variability of the imputed data and can be severely biased \citep{14dempster1977maximum}. For these reasons we will not be pursuing Single Imputation in this paper but instead show the usefulness of Multiple Imputation (MI) \citep{15rubin1987multiple}, where the multiple imputed  datasets are analysed using standard statistical procedures and the estimates from these analyses are then combined to produce overall estimates with more appropriate standard errors. These estimates and standard errors reflect the true variation and uncertainty in the data better than the estimates obtained by deleting any observations with missing values, also known as available case analysis. MI is less biased than single imputation methods, increases the efficiency of estimation and also preserves the variability in the dataset.  Hence, it is the preferred method to fill in missing values in datasets such as the one used in this study. It should be noted that the goal of imputation is not to replace or recover these missing values from the dataset, as they are unknown, but to produce valid and analytical results in the presence of the missing values.\\
\\
Despite its conception in the 1980s, MI has only recently come into general use across medical research due to advances in statistical computation.  Even with the evolution of MI and its  statistical advantages, the technique has had limited application in laboratory diagnostics. The use of MI in clinical chemistry has been supported \citep{16janssen2009dealing, 17waljee2013comparison} when values of a variable in an existing prediction model are missing.The study presented herein extends this previous approach by imputing variables with an unknown relationship to the outcome of interest. \\
\\
The primary goal of this study is to show the importance of multiple imputation when dealing with missing values in routinely collected pathology data. A secondary goal of this paper is to contribute to knowledge of laboratory prediction of HCV by identifying novel combinations of routinely measured biomarkers that are highly predictive of HCV. This focus on an existing resource which provides a low-cost approach to knowledge discovery is important in the context of low-middle income countries, where laboratory resources may be limited \citep{18shang2013predicting}.

\section{Materials and Methods}
\subsection{Data}
The dataset employed in this study was made available by ACT Pathology at The Canberra Hospital, Canberra, Australia. Patient identifiers were removed and only laboratory ID numbers provided. The data set contained the 18,625 pathology requests between 1997 and 2007 that included a re-quest for either the assay for Hepatitis B (HepB) or Hepatitis C (HepC) or both (Figure 1). There were 4,296 patients for whom 60\% or more of the other assays (Table 1) were missing, and these were removed before analysis commenced due to the large proportion of missingness. Another 3,546 individuals were missing a HCV immunoassay (HepC) outcome despite a request having been made and so excluded from the analysis. The result of the HCV immunoassay is recorded as positive if the test is positive to the presence of HCV anti-bodies. The same dataset has been analysed previously \citep{2richardson2013infection, 3richardson2017enhancement} to ascertain the interaction between virus, outcome, pre-processing and method on the performance of decision tree ensembles and support vector machines.\\
\\
For access to de-identified patient data, this study had human ethics approval granted by The Australian National University Human Ethics Committee (2012/349) and the ACT Health Human Research Ethics Committee (ETHLR.11.016).

\begin{figure}[H]
\includegraphics[width = \linewidth, height = 4.5in ]{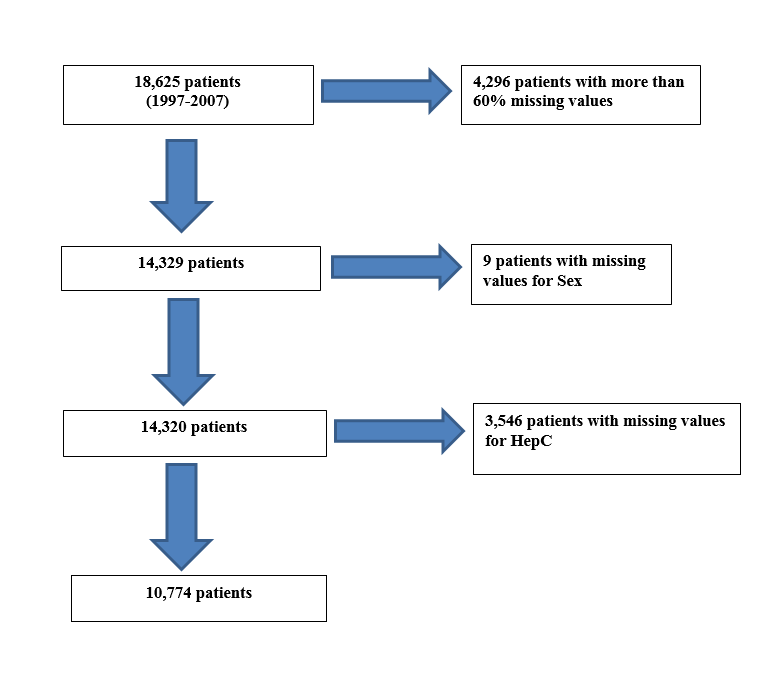}
\caption{Flow of patients into the HepC study \textit{(To be printed in Colour)}}
\label{Fig 1}
\end{figure}

\begin{table}
\caption{Diagnostic pathology variables included in the regression analysis, with summary statistics (n = 10,774).}
\begin{flushleft}
\begin{tabular}{p{1.5cm} |p{5.2cm}| p{2.5 cm}|p{2.5cm}| p{2cm}}
\hline
\textit{Variable}& \textit{Description} & \makecell{\textit{Mean (SD)}\\\textit{ (HepC = 0)} \\\textit{(n = 10102)}} & \makecell{\textit{Mean (SD)}\\ \textit{(HepC = 1) }\\\textit{(n = 672)}} & \textit{ p-value}\\
\hline
Age & Patient’s age in years & $44.72 \ (19.36)$ & $40.09\ (14.41)$ & $<0.001$\\
Sex & \makecell[l]{Patient’s gender (Males) \\Frequency (\%)} &  $5188\ (51.36 \%)	$ & $251\ (37.35 \%)	$ & 0.00013\\
ALB & Albumin & $42.84 \ (5.80)$ & $42.09\  (5.88)$ & 0.0004\\
Sodium		&	Sodium								& $139.63\   (3.22)$	& $139.66 \  (3.06)$ 		&	$0.9927$\\
K	&	Potassium							&		$ 4.00 \  (0.45)$ 	&		$4.07 \ (0.47)$		&	$<0.001$\\
RCC			&	Red cell count 						&$4.53 \ (0.64)$	&	$4.68\  (0.64)	$			&	$<0.001$\\
RDW			&	Red cell distribution width &$13.88 \  (1.78)$	&	$14.07\  (1.63)$			&	$<0.001$\\
ALT			&	Alanine aminotransferase &	 	$1.46 \  (0.38)$&	1$.64       \  (0.42)$		&	$<0.001$\\
ALKP			&	Alkaline Phosphate		 &		$1.92\  (0.21)$	&	1$.94       \  (0.18)$		&		$<0.001$\\
Crea			&	Creatinine						 &		$1.93  \  (0.16)$&		$1.90       \  (0.13)$	&	$<0.001$\\
GGT			&	Gamma-glutamyl transferase &		$1.61  \  (0.44)$	&	$1.69       \  (0.43)$	&		$<0.001$\\
Urea			&	Blood urea						&		$0.72 \  (0.20)$	&	$0.65       \  (0.19)$		&	$<0.001$\\
Plt				&	Platelets								&	$2.39 \  (0.20)$		&		$2.39       \  (0.18)$	&$0.2179$\\
WCC			&	White cell count				&		$0.87 \  (0.17)$	&		$0.90       \  (0.15)$	&		$<0.001$\\
TBil				&	Total bilirubin					&		$1.03 \  (0.30)$		&	$0.98       \  (0.31)$		&		$<0.001$\\
Mono			&		Monocytes						&	$0.74  \  (0.20)$		&		$0.76       \  (0.15)$	&		$<0.001$\\
Eos				&		Eosinophils					&		$0.38   \  (0.18)$	&		$0.39       \  (0.18)$	&		$0.014$\\
Bas				&	Basophil								&		$0.18 \  (0.09)$	&		$0.19       \  (0.08)$	&	$<0.001$\\
Lymph		&	Lymphocytes						&	$1.38\  (0.34)$			&	$1.48       \  (0.53)$		&		$<0.001$\\
\end{tabular}
\end{flushleft}
\end{table}

\subsection{Statistical Analysis}
This is a cross-sectional dataset where each individual only appears once with a non-monotone pattern of missingness. Percentage missingness is calculated for each variable and we examine the pattern of missingness visually to get an impression of the extent of incompleteness. These patterns of missingness influences the amount of information that can be transferred between variables \citep{19van2012flexible}. For example, imputations can be more precise if complete information is available for the other variables for the observations that are to be imputed. \\
\\
In his book \citep{19van2012flexible}, van Buuren proposed using Influx and Outflux statistics to streamline this process  of constructing imputation models that are impartial to the subjectivity of the imputer and the analyst \cite{19van2012flexible}. These statistics quantity how each variable connects to others. 
As described by \cite{19van2012flexible}, for a pair of variables $(X_a, X_b)$ in a sample of $n$ observations with $p$ variables, the influx coefficient for the variable $X_a$ is defined as $$ I_a = \frac{\sum_{a =1}^p\sum_{b=1}^p\sum_{i=1}^n (1-r_{i,a}) r_{i,b}}{\sum_{b = 1}^p \sum_{i=1}^nr_{i,b}}$$
The value of $I_a$ depends on the proportion of missing data in the variable $X_a$. This means that if $X_a$ has no missing values, $I_a$ = 1. The Influx coefficient ($I_a$) is the number of variable pairs $(X_a, X_b)$, where $X_a$ is missing and $X_b$ is observed. Thus, a high influx implies how well the variable is connected with the observed data. \\
\\
The second measure proposed by van Buuren, is the outflux coefficient. For the same pair of variables $(X_a, X_b)$, the outflux coefficient indicated how useful is $X_a$ to impute missing values in $X_b$. Unlike the former, outflux depends on the extent of missing values in the variable $X_a$. The variable with higher outflux is better connected to the missing data, and can be more useful in imputing other variables \citep{19van2012flexible}. 
The outflux coefficient is given by $$ O_a = \frac{\sum_{a=1}^p\sum_{b=1}^p\sum_{i=1}^n  r_{i,a}(1-r_{i,b})}{\sum_{b=1}^p \sum_{i=1}^n(1-r_{i,a})}$$ \\  
\textit{Fluxplots} can be used to plot both these measures to make interpretation easier for the imputer and the analyst. It can be used to identify variables that clutter the imputation model, thus making the process of constructing the imputation model less subjective.  Variables in the lower areas in the fluxplot that are not used for analysis can be removed from the data prior to imputation. While there is a significant proportion of literature that focuses on developing proper imputation models, there is an evident lack of illustrative examples describing the imputation process in practice to create imputation models are as impartial to the imputer and the analysts and relies on quantitative measures to develop these models.
\\
Multiple imputation is used to address missing data. In this method, each missing value is replaced by two or more imputed values to represent the statistical uncertainty about which value to impute. It involves three stages; Imputation, Analysis and Pooling. Final inference on the regression coefficient estimates is made on the pooled result using Rubin’s combination rule \citep{13little2002statistical}.  Variables with an Outflux coefficient above 0.95 i.e. variables that occurred in the top left corner of the flux plot, are used as predictors in the imputation model. We employ a stepwise procedure by calculating the number of times a variable occurred a predictor in the imputation model. This method is also referred to as the impute then select method \citep{20yang2005imputation}. A supermodel is constructed using the variables that occurred across all imputed datasets. For all other variables, we use the multivariate Wald test and the likelihood ratio test to determine whether the variable should be included in the final model. \\
\\
Multiple Imputation using Chained Equations (MICE) \citep{19van2012flexible} has emerged in the literature as a routine method for handling missing data in both continuous and binary variables. Descriptive statistics (mean and standard deviation) for all iterations are used to obtain a plot to check imputation convergence. If the mean and standard deviation of the imputed variables appear to have settled at particular values, the imputation process is deemed to have converged. There is a lack of formal tests of convergence \citep{21su2011multiple}, so the plots of statistics such as the mean and standard deviation are used to provide visual information about convergence. A backwards step wise model selection procedure is employed in the analysis phase after imputations. \\
\\
All analysis was performed using the statistical software R (version 3.4.0) \citep{22}.  The package ‘mice’ \citep{23van2011mice} were used for multiple imputation. The package ‘ROCR’ \citep{24sing2005rocr} is used to produce Receiver Operating Characteristic (ROC) curves to test the predictive ability of the imputed models and to calculate the area under the curve (AUC).  The packages ‘nortest’ \citep{25gross2015package} and ‘psych’ \citep{26} are used to perform the Anderson-Darling test of normality and to obtain summary statistics respectively. The package ‘MASS’ \citep{27venables2002modern} is used to perform $\chi^2$ and Mann-Whitney U tests.   \\

\begin{table}[H]
\caption{Diagnostic pathology variables included in the regression analysis, with summary statistics (n = 10,774).}
\begin{flushleft}
\begin{tabular}{p{3cm} | p{4cm}|p{4cm}}
\hline
\textit{Variable} & \makecell{\textit{Percentage Missing}\\\textit{ (HepC = 0)} \\\textit{(n = 10102)}} & \makecell{\textit{Percentage Missing}\\ \textit{(HepC = 1) }\\\textit{(n = 672)}}\\
\hline
Age &	0 &		0\\
Sex &	0 &	0\\
Urea	 &	18.71	 &	13.39\\
TBil	 &	18.71	 &	4.46\\
ALT	 &	18.67 &		4.46\\
Crea	 &	18.65 &		13.39\\
GGT	 &	18.63	 &	4.02\\
Potassium	 &	17.52	 &	12.35\\
ALKP &		17.43	 &	3.57\\
Sodium	 &	17.19 &		11.61\\
ALB &		16.41	 &	3.57\\
Bas &		2.52 &		2.23\\
Eos	 &	1.97 &	1.04\\
Plt	 &	0.86 &		0.15\\
Mono	 &	0.53 &		0.30\\
Neut	 &	0.53 &		0.30\\
Lymph	 &	0.53	 &	0.30\\
WCC	 &	0.50 &	0\\
MCHC	 &	0.23 &		0\\
Hct &		0.21	 &	0\\
RCC	 &	0.21 &		0\\
RDW	 &	0.17	 &	0.45\\
MCV	 &	0.16	 &	0\\
Mch &		0.16 &		0\\
Hb &		0.12 &		0\\
\end{tabular}
\end{flushleft}
\end{table}

\begin{figure}[H]
\includegraphics[width = \linewidth, height = 5in ]{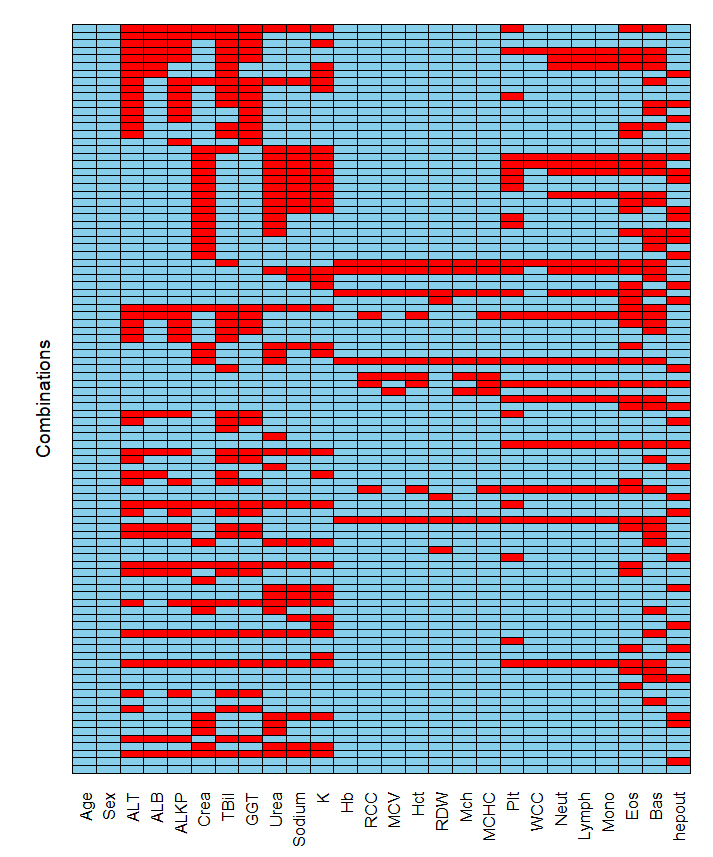}
\caption{Patterns of missingness in predictor variables by combination. Each row of the chart represents a different combination of missing values.\\ Blue = a variable has no missing values, red = a variable has missing values. See Table 2 for abbreviation of biomarker names.\textit{(To be printed in Colour)} 
}
\label{Fig 2}
\end{figure}

\begin{figure}[H]
\includegraphics[width = \linewidth, height = 4in ]{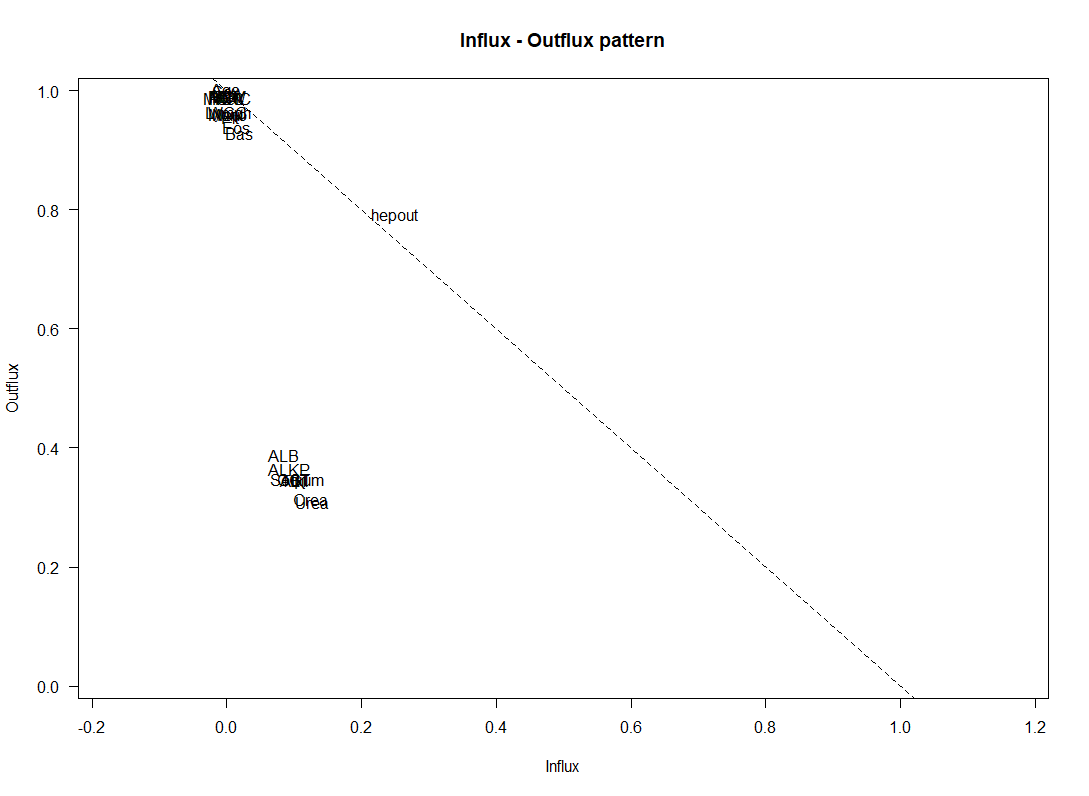}
\caption{Fluxplot for HepC data: Outflux v/s Influx}
\label{Fig 3}
\end{figure}

\begin{table}[H]
\caption{Influx and outflux of multivariate missing data patterns.   }
\begin{flushleft}
\begin{tabular}{p{2cm} |p{2cm}| p{2cm} |p{2cm}| p{2cm}}
\hline
\textit{ Variable} &    \textit{\makecell{Proportion\\Observed}}   &    \textit{Influx}  &\textit{Outflux }&   \textit{FICO}\\
\hline
Age  &  1.000   &   0.000   &    1.000   &  0.454\\
Sex     &   1.000   &   0.000   &    1.000   &  0.454\\
ALT    &    0.865   &   0.101   &    0.346   &  0.369\\
ALB     &   0.882   &   0.086    &   0.387   &   0.381\\
ALKP     &  0.875    &  0.093    &   0.365    &  0.376\\
Crea    &   0.840   &   0.126    &   0.314    &  0.350\\
TBil    &   0.865   &   0.102    &   0.345    &  0.369\\
GGT     &   0.866   &   0.101    &   0.348    &  0.369\\
Urea    &   0.839   &   0.127    &   0.310    &  0.349\\
Sodium   &  0.862   &   0.105    &   0.348    &  0.366\\
K       &   0.858   &   0.109    &   0.343  &   0.364\\
Hb      &   0.999    &  0.000   &    0.992   &   0.453\\
RCC    &    0.998    &  0.001    &   0.987  &   0.453\\
MCV    &    0.999   &   0.001    &   0.992   &   0.453\\
Hct    &    0.998    &  0.001    &   0.987   &   0.453\\
RDW    &    0.998    &  0.001   &    0.991   &   0.453\\
Mch    &    0.999    &  0.001    &   0.992   &   0.453\\
MCHC   &    0.998   &   0.001    &   0.987  &   0.453\\
Plt     &   0.993   &   0.005    &   0.956   &  0.450\\
WCC    &    0.996    &  0.002   &    0.964   &  0.452\\
Neut    &   0.996    &  0.002    &   0.961   &   0.451\\
Lymph   &   0.996    &  0.002    &   0.961   &  0.451\\
Mono    &   0.996    &  0.002    &   0.961   &   0.451\\
Eos    &    0.983    &  0.014    &   0.939   &   0.444\\
Bas    &    0.978    &  0.019    &   0.928   &  0.442\\
hepout  &   0.752    &  0.250    &   0.791  &   0.274\\
\end{tabular}
\end{flushleft}
\end{table}

The variables are imputed on the raw scale. Log or square root transformations are applied to those variables with highly skewed distributions using passive imputation \citep{28von20074}. Passive imputation is a method to handle derived variables in imputation by transforming the imputed values of the variable. Even though a Normal distribution is not essential in the predictors, transformations were applied to ensure predictors were of similar order of magnitude across analysis methods. Complex transformations (like Box-Cox power transformations) were not considered for the final model for clarity of interpretation. Finally, logistic regression is used to predict HCV status on the basis of age, sex and the 23 biomarkers. \\
\\

\section{Results}
There were over 100 different patterns of missingness in the predictor variables across the 14,320 patients (Figure 2a). There were 7,820 individuals (54.6\%) with complete data on age, sex and all 23 biomarkers. 
\\
Percentages of missingness (Figure 2b, Table 2) occurs in one of three ways. For Sodium, Potassium, creatinine (Crea) and Urea, there is approximately 15\% missingness, whether or not a patient was HCV positive. For serum albumin (ALB), alanine aminotransferase (ALT), alkaline phosphatase (ALP), gamma glutamyl transferase (GGT) and total serum bilirubin (TBil), there is about 15\% missingness amongst HepC negative, and less than 1\% missingness amongst HepC positive. For the remainder (Age, Sex, platelets (Plt), monocytes (Mono), eosinophils (Eos), basophils (Bas), lymphocytes (Lymph)), there was only occasional missingness of data.\\
\\
The area under the ROC curve (AUROC) (Figure 5a) for the available cases logistic regression model on the validation dataset was 72\%; namely, the model has a 72\% chance of correctly classifying individuals as having a positive HepC assay. The model identified the variables Age, Sex, Sodium, RDW, Potassium, log(ALT), log(ALP), log(Crea), log(TBil), log(Urea) and sqrt(Lymph) as significant $(p < 0.05$; see Table 3).\\
\\
As stated earlier, variables with higher outflux are potentially powerful predictors and can be used to impute variables with missing values. As indicated by the flux plot (Figure 3), variables in the far left corner have more complete data. Variables closer to the diagonal have more balanced values of influx and outflux. The group below the diagonal with higher values of influx depend highly on the imputation model. In this study, we consider variables with outflux values $> 0.90$ to be included in the imputation model. These comprise: Age, Sex, Hb, RCC, MCV, Hct, RDW, Mch, MCHC, Plt ,WCC, Mono, Eos, Bas, Neut and  {Lymph}. \\
\\
We fit a stepwise logistic model to predict HepC and count the number of times each variable occurred in the imputation model in each of the imputed datasets. This was done for five, and 20 imputations.  We observe that variables that did not occur in any of the imputation models when there were five imputations, occur in lower frequencies when there were 20 imputations. 

\begin{figure}[H]
\includegraphics[width = 6in, height = 4.5in ]{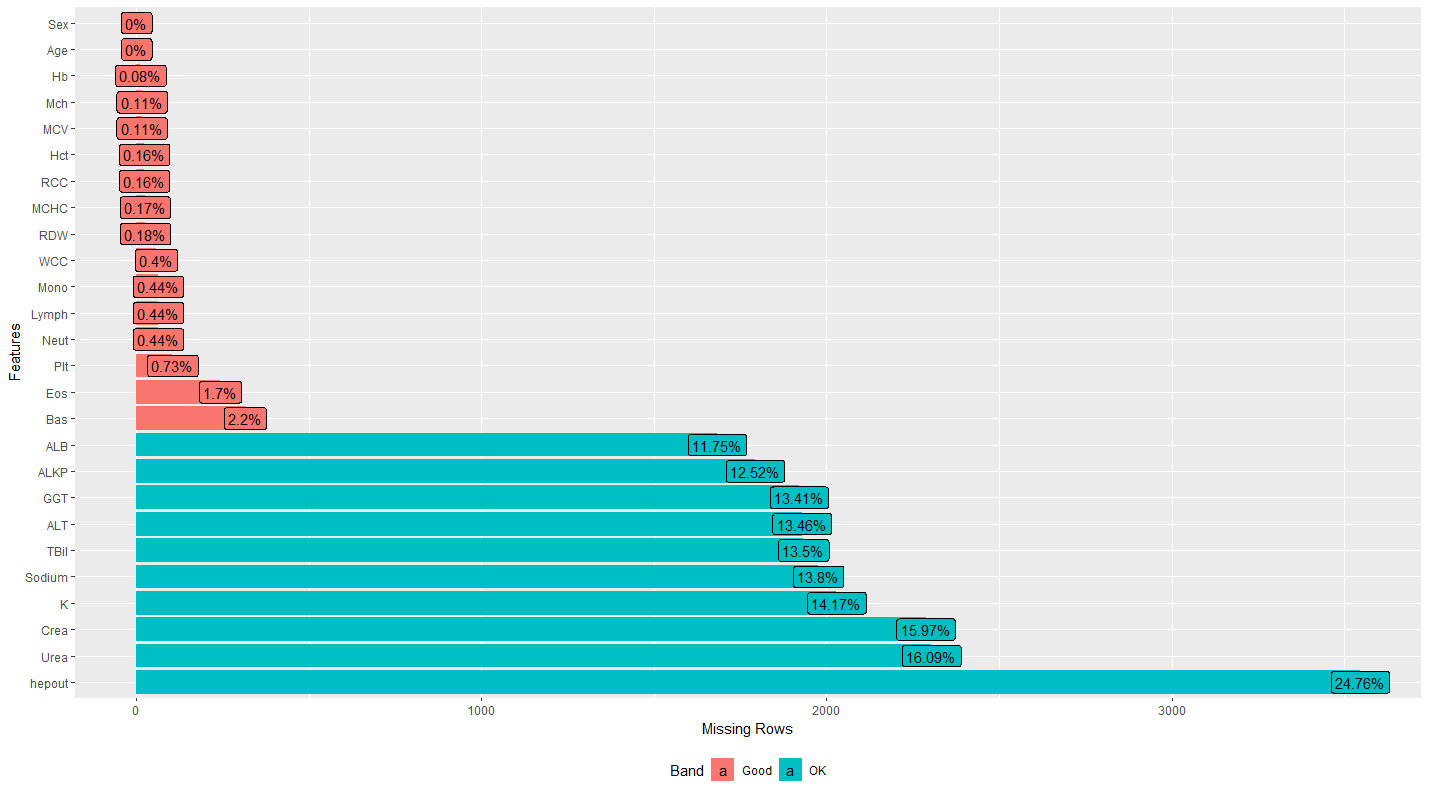}
\caption{Percentage of missingness across variables in the dataset. Variables coloured pink have between 0 and 10\% mising values, variables in blue have between 11 and 25\% missing values. \textit{(To be printed in Colour)}}
\label{Fig 4}
\end{figure}

\begin{table}[H]
\caption{Tabulation of the number of occurrences in imputation model across all imputed datasets.   }
\begin{flushleft}
\begin{tabular}{p{3cm} |p{4cm}| p{4 cm}}
\hline
\textit{Variable}&  \makecell{\textit{No. of Occurrences }\\\textit{(5 imputations)}} & \makecell{\textit{No. of Occurrences }\\\textit{(20 imputations)}}\\
\hline
Age &	5	&	20\\
ALB	&	5	&	20\\
ALT&		5	&	20\\
Lymphocytes&		5	&	20\\
RCC	&	5	&	20\\
RDW&		5&		20\\
Sex	&	5	&	20\\
TBil&		5	&	20\\
Urea&		4&		20\\
Crea	&	5	&	18\\
K	&	3	&	15\\
MCV	&	4	&	14\\
Basophils &		5	&	13\\
Hb	&	3	&	13\\
MCHC &		3	&	13\\
Sodium &		4	&	13\\
Hct	&	2	&	9\\
ALKP	&	0 &		7\\
Monocytes	&	2&		6\\
Eosinophils	&	0	&	5\\
Mch	&	3	&	5\\
WCC	&	0&		3\\
Neut	&	0	&	2\\
Plt	&	0	&	1\\
\end{tabular}
\end{flushleft}
\end{table}

Variables that occurred in all imputation models were considered in the supermodel.For other variables, the likelihood ratio test was applied to identify if the variable should be included in the final model. Variables with p-value $< 0.05$ were selected in the final supermodel. 
Table 4 shows the results of the tabulation for number of occurrences of each variable when number of imputations were equal to 5 and 20.

\begin{landscape}
\begin{table}
\caption{Regression coefficients for logistic regression models. Log = log transform applied before regression analysis. Sqrt = square root transformation applied before regression analysis }
\begin{flushleft}
\begin{tabular}{p{2cm} |p{3cm}| p{3 cm} | p{3cm}| p{3cm}|p{3cm}| p{3 cm}}
\hline
%\textit{Variable}&  \makecell{\textit{Available Case Analysis}\\\textit{(n = 7,820)}} & \makecell{\textit{Multiple Imputation (m = 5) }\\\textit{(n = 14,320)}} &  \makecell{\textit{Multiple Imputation (m = 20) }\\\textit{(n = 14,320)}}\\
\textit{Variable}&  CC: $\beta (SE)$ & CC: p-value &  MICE (m = 5): $\beta (SE)$ & MICE (m = 5):  p-value & MICE (m = 20):  $\beta (SE)$ & MICE (m = 20):  p-value \\
\hline
Intercept & 3.408 (2.455)	& 0.1651 & 	31.577 (8.937)	& $<0.001$	& -0.7547 (3.2853) & 	0.8183\\
Age &	-0.016 (0.003)&	$<0.001$	&-0.016 (0.003)	&$<0.001$&	-0.0169 (0.003)&	$<0.001$\\
Sex	&0.664   (0.109)	&$<0.001$&	0.516 (0.112)&	$<0.001$	&0.5043 (0.0956)&	$<0.001$\\
RDW	&0.087     (0.026)	&$<0.001$	&0.142 (0.023)	&$ <0.001$ &	0.1592 (0.0254)&	$<0.001$\\
Log(ALT)	&1.415   (0.123)	&$<0.001$	&0.3504 (0.0481)	&$<0.001$ &	0.3436 (0.0445)&	$<0.001$\\
Sqrt(Lymph)	& 0.312 (0.112)	& 0.0056	& 0.3912 (0.099)	& $<0.001$	& 0.3406 (0.0943) & 	$<0.001$\\
ALB	&-0.016    (0.009)&	0.0644	&-0.044 (0.008)&	$<0.001$ &	-0.0473 (0.0084)	& $<0.001$\\
Potassium	&0.383     (0.112)&	$<0.001$	& -- &  -- &	0.1439 (0.0962)	&0.1346\\
Sodium	&-0.035    (0.016)	&0.0292	& --	& -- & --	& --\\
Sqrt(Bas) & 	0.701    (0.361)	& 0.0523	 & -- & 	-- &  -- & 	-- \\
Log(ALKP)& 	-0.921    (0.269)	& $<0.001$ &	--  & 	-- &	-- & 	--\\
Log(Crea)	&-1.320    (0.558)&	0.0180&	-0.3904 (0.195) & 	0.0046 & 	-- & 	--\\
Log(TBil)	&-1.114    (0.170)&	$<0.001$	& -0.3945 (0.065)& $	<0.001$	& -0.4031 (0.0656)& $	<0.001$\\
Log(Urea)	&-1.261  (0.338)	& $<0.001$&	-0.251 (0.135)& 	0.0652	& -0.4706 (0.114)& 	$<0.001$\\
Sqrt(Mono)	& --	& -- & -0.174(0.275)	& 0.5251 & 	-- & --\\
Hb	&-- & -- &	0.0913(0.022) & $	<0.001$ &	0.0774(0.0335)& 	0.0211\\
MCHC	&-- & -- &	-0.0902(0.024) & $	<0.001 $		& -- & -- \\
MCV	&-- & -- & -0.2328 (0.0889)&	0.0089	& 0.0712 (0.0549)	& 0.0194\\
RCC	&-- & -- & -2.107 (0.698)	& 0.0025 & 	-2.0800 (0.7416) & 	0.0051\\
Mch	&-- & --  & 	0.503 (0.2762)	& 0.0688	 & -0.4026 (0.1620) & 	0.0131\\
Hct	&-- & --  & -- & -- &	4.3233(11.205)	& 0.0699\\
\end{tabular}
\end{flushleft}
\end{table}
\end{landscape}

\begin{landscape}
\begin{figure}
  \label{fig}\caption{Receiver-Operator Characteristic (ROC) curves for each method of analysis. Diagonal line represents ROC of 50\% \textit{(To be printed in Colour)}}
  \subfloat[Available Case analysis]{\label{ROC_cc}\includegraphics[width=95mm]{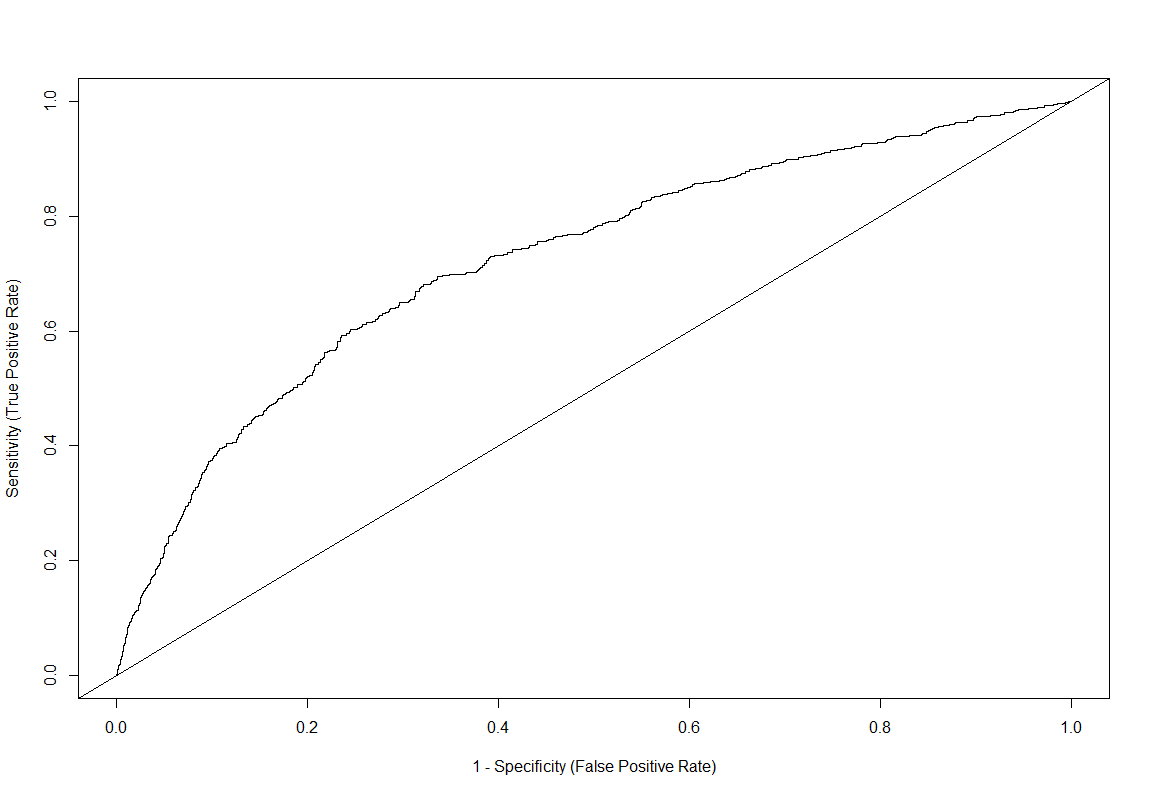}}
  \\
  \subfloat[MICE (m = 5)]{\label{figur:2}\includegraphics[width=95mm]{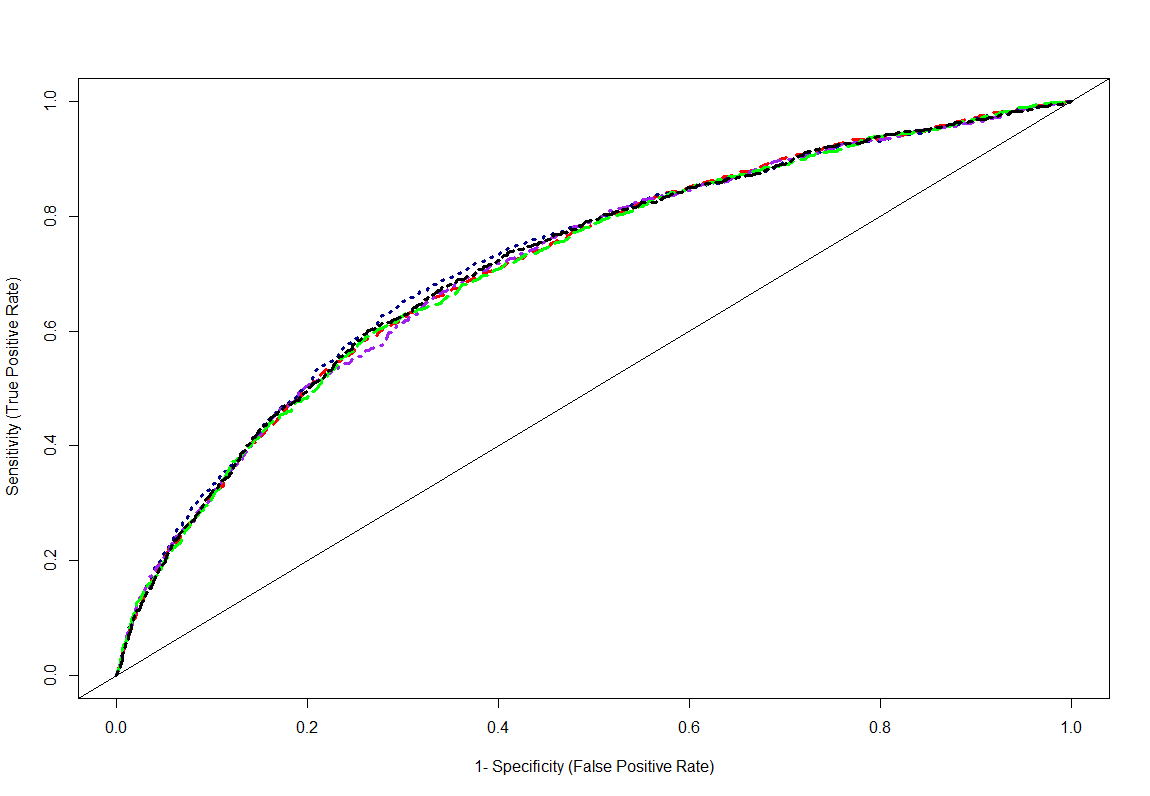}}
  \subfloat[MICE ( m = 20)]{\label{figur:3}\includegraphics[width=95mm]{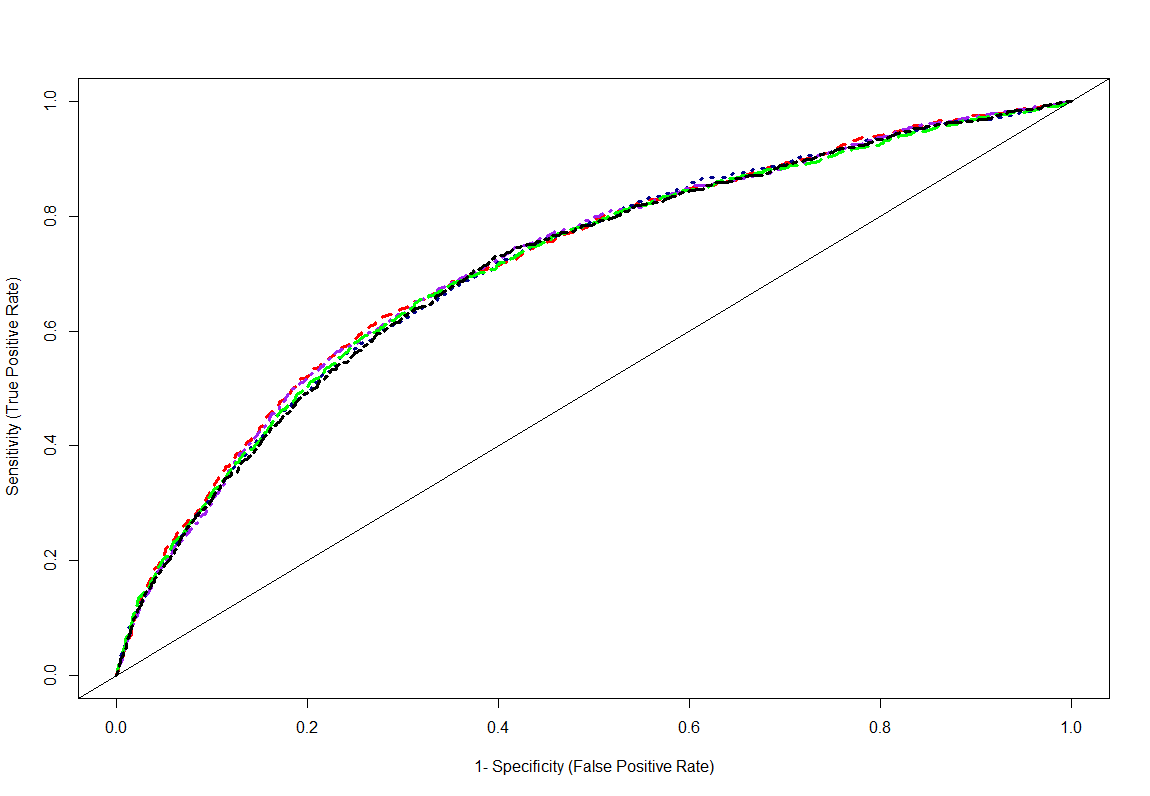}}
\end{figure}
\end{landscape}

The parameter estimate from the multiple imputation methods are automatically pooled and presented in Table 5. The AUROCs of the five-fold imputed models have a mean of 72\% (SD 0.003). We increased the number of imputations from 5 to 20 and observed variable selection was more consistent, reducing the randomness in the variable selection process. The AUROCs of the twenty-fold imputed models have a mean of 71\% (SD  0.002). 

\section{Discussion}

Discussion of the results will involve both the meaning of the results for multiple imputation in general, and the results for the diagnosis of hepatitis C in particular. The focus of the study is to improve the diagnostic process by developing a powerful predictive model for HepC without discarding any information collected during laboratory tests due to the presence of missing values, and using the completed dataset for developing this model. \\
\\
The percentage of missing values displayed three patterns (see Results). Biomarkers with around 15\% missingness included sodium, potassium, creatinine and urea which are associated with the Urea-Electrolytes-Creatinine battery of tests. These are routinely requested for most laboratory investigations. Biomarkers such as ALB, ALT, ALP, GGT and TBil and displayed around 15\% missingness for those found HCV negative, and less than 1\% missingness for those found HCV positive. These are routine liver function tests and would be primary markers sought when HepC is suspected. The remaining biomarkers are from the full blood count and are routinely requested, and so records are complete for almost all individuals. A recent paper \citep{29lidbury2018predicting} has been published on this strategy, but with the goal of avoiding painful and expensive liver biopsy. \\
\\
The multiply imputed analyses are largely consistent with the available case analysis \citep{2richardson2013infection, 3richardson2017enhancement} (Table 3).  However, three variables in the single imputation model were not identified as important disease markers in the analysis of the imputed datasets; hence some unnecessary variables were removed via MI. Although the ROC curves (Figure 5) produced by the MI methods did not yield high absolute accuracy, the ten key predictors common to all the multiply imputed regression models suggest a pattern of routine, easily accessible pathology markers, which predict a HCV infection. In addition to early detection, the advantage of using quality routine pathology data as a predictive model, where immunoassay is not easily available is a key outcome especially when handling population health issues for rural or remote areas in third world countries.\\
\\
The regression models suggest that increased levels of ALT, RDW and lymphocytes increased the odds of detecting HCV infection. All models presented include the variable ALT which implies that ALT is one of the most useful routine test predictors of HCV infection. The finding for ALT coincides with current medical practice \citep{30holmes2013hepatitis}, as cases with elevated ALT levels are more likely to be infected with HCV. However, ALT elevation is not specific for viral hepatitis [\citep{31kwo2017acg} and so an increased level of ALT in the blood is used as a guide to request specific second-tier tests that may include HepC immunoassay.\\
\\
RDW has also been previously identified as having a relationship with HCV in a small-scale Chinese in-patient study \citep{18shang2013predicting}; the current analysis supports that finding in a much larger Australian community cohort. Lymphocytes have also been identified as playing a role in HCV infection in a systematic review of epidemiological studies \citep{32he2016relationship}. \\
\\
Age is another strong predictor of HCV status, with increasing age decreasing the chance of being HCV positive.  Evidence from Europe \citep{33cacoub2016extrahepatic} and Australia \citep{34faustini2010hepatic} supports this. The increased prevalence in this study of HCV diagnosis in younger patients suggests that clinicians could use this evidence to be proactive in seeking information on the key variables identified in this study in the younger age groups. Although the laboratory supplying the data is housed on a hospital campus, the organisation provides pathology services to the public as well as inpatients, hence the sample contains many community-living persons with a wide range of clinical indications leading to the request for an HCV immunoassay.  \\
\\
The multiple imputation methods employed here assume MAR. However, the results of this study are limited by the likelihood that MNAR is actually the type of missingness observed in pathology data. Policy initiatives such as those recently employed for Vitamin D \citep{35boyages2016vitamin} provide external reasons to reduce the number of pathology tests ordered in a variety of situations. Another likely reason that MNAR would arise in the context of this study is that clinicians are likely to use their clinical judgement to order some tests and not others. This clinical judgment is now also being enhanced by data mining \citep{36lidbury2015assessment}. Non-clinical characteristics can also drive choice of tests requested \citep{37smellie2002clinical}. Practice characteristics and other clinical notes are not captured in the available data. \\
\\
A second and related limitation of this multiple imputation analysis is that were MAR to be assumed, the parameter estimates from the imputed data are very sensitive to the model employed for the probability of response \citep{13little2002statistical}. The probability of response can be modelled by using linear regression models for continuous variables, logistic regression models for binary variables, and multinomial logistic regression models for categorical variables with more than two classes.\\
\\
A final limitation of this analysis concerns the rarity of the outcome: only 3\% of subjects tested positive for HCV. Whilst the rarity of the outcome supports our view that inclusion bias is likely to be small, it does raise issues around the stability of classical estimation algorithms. Logistic regression for rare events may require the use of penalised likelihood instead of maximum likelihood \citep{38king2001logistic}.

\section {Conclusion}

Faced with missing data in a routine pathology laboratory database, logistic regression to develop a prediction model for HCV positive assay results was successfully undertaken following the use of multiple imputation.  The coefficient estimates of the regression models built on various imputed datasets were pooled to obtain overall estimates.  These logistic regression models, on average, explained about 70\% of the variation in the imputed dataset.  The regression models built on imputed datasets have similar AUROCs compared to the regression model generated using the complete dataset (73\% on complete dataset compared to an average of 71\% to 72\% on multiply imputed datasets).  However, the regression models on the imputed datasets identified more predictors as significant in predicting HCV infection. This means that MI leads to richer prediction models and not using MI may lead to missing predictors that have a useful relationship with the outcome of interest. This study suggests that MI in the diagnostic process of HepC infection using laboratory data has identified combinations of routinely measured biomarkers that are integral to the prediction of HCV infection. \\
\\
\textbf{Acknowledgements:} The authors wish to thank Dr Gus Koerbin and staff at ACT Pathology for their support of this project. They also thank Mr Sheikh Faisal who undertook the initial analysis of this data as part of his Master’s degree research.\\
\\
\textbf{Financial support:} This research did not receive any specific grant from funding agencies in the public, commercial, or not-for-profit sectors.\\ 

	%%%%%%%%%%%%%%%%%%%%%%%%%%%%%%%%%%%%%%%%
	%\bibliographystyle{CBS}
	\bibliographystyle{chicago}
	\bibliography{hepC_2022_updated}
	%%%%%%%%%%%%%%%%%%%%%%%%%%%%%%%%%%%%%%%%

\end{document}